\documentclass{scrartcl}
\usepackage{graphicx} 

\usepackage{amsmath}

\usepackage{siunitx}
\begin{document}

\date{}
\title{Spin Wave Threshold Logic Gates}

\author{Arne Van Zegbroeck, Pantazis Anagnostou, Said Hamdioui, \\ Christop Adelmann, Florin Ciubotaru, Sorin Cotofana}

\begin{abstract}
    While Spin Waves (SW) interaction provides natural support for low power Majority (MAJ) gate implementations many hurdles still exists on the road towards the  realization of practically relevant SW circuits. In this paper we leave the SW interaction avenue and propose Threshold Logic (TL) inspired  SW computing, which relies on successive phase rotations applied to one single SW instead of on the interference of an odd number of SWs. After providing a short TL inside we  introduce the SW TL gate  concept and discuss the way to mirror TL gate weight and threshold values into physical phase-shifter parameters. Subsequently, we design and demonstrate proper operation of a SW TL based Full Adder (FA) by means of micro-magnetic simulations. We conclude the paper by providing  inside on the potential advantages of our proposal by means of a conceptual comparison of MAJ and TL based FA implementations.
    
\end{abstract}







\maketitle

\section{Introduction}
The amount of data that is processed every day has significantly increased  over the past decade \cite{BigData}. However, the computing devices efficiency and power consumption has not scaled equally with the increase of data causing a big jump in the global IT power consumption in that time period \cite{EnergyCons}. This is even aggravated by the fact that transistor scaling is running into more and more problems due to short channel effects, power density and gate tunneling, to name a few \cite{CMOSscaling}. To address this issues, new FET architectures like FinFET have been proposed \cite{FinFetScaling} and technologies that completely depart from CMOS, commonly referred to as beyond-CMOS, are under scrutiny, as they can potentially enable new avenues for power efficient data processing. Examples include, but are not limited to, Graphene  \cite{GrapheneComp,GNRpaper}, Quantum  \cite{Quantum}, and Spintronics  \cite{ReviewSpin, NanoMagnet}. 

Among those, Spintronics, which makes use of electron spin for information encoding,  provides powerful means for the implementation of low power circuits, efficient non-volatile memories \cite{SpintronicsReview}, and neuromorphic circuits \cite{SpintronicsNeuromorph}. Recently, Spin Waves (SW), which are small collective magnetization deviations that travel in a wave like manner \cite{Chumak2019FundamentalsOM} through a magnetised material, received special attention as their high frequency and small wavelength provide premises for fast and small circuit implementations \cite{Chumak2019FundamentalsOM, IntroSpinComp}. 


Current research on SW logic mostly revolves around the implementation of classic Boolean functions \cite{IntroSpinComp}, by encoding binary data in SW phase and letting an odd number of unit amplitude SWs interfere within a common waveguide.  Due to the very nature of the SW interference process it provides natural support for majority function evaluation.  Essentially speaking, related to a certain reference, input SW are either in phase (logic $0$) or $\pi^{\circ}$ out of phase (logic $1$). Within the waveguide in-phase/out-of-phase SWs constructively/distinctively interact resulting in a SW having the phase of the majority of the SW inputs. The majority value can be obtained in its direct and/or inverted form by properly adjusting the output SW amplitude reading position \cite{SWFanout}. By following this concept $3$-input majority gates ($MAJ3$) have been proposed and simulated and/or experimentally demonstrated \cite{SpinwaveMajority, InlineMajorityGate, SWFanout}. As $MAJ3$ and inverter form a universal gate set any Boolean function can be implemented by following this paradigm.


In this paper we leave the SW interaction avenue and propose Threshold Logic (TL) inspired SW computing. To implement an $n$-input TL gate, instead of inducing multiple SWs and letting them interact within a waveguide, we make use of one single SW on which we induce $n+1$ successive phase rotations. The final SW phase sign carries the gate output value, i.e., negative logic $0$ and positive logic $1$. We introduce this novel concept, provide inside on the SW TL gate (SWTLG) design methodology, and provide preliminary inside on the potential impact of our proposal.

The paper is organized as follows: In Section 2 we briefly present Threshold Logic (TL) fundamentals. In Section 3 we introduce the novel concept of SW phase manipulation based computing  and in Section 4 provide inside on SW phase manipulation and the SW TL gate design process. In Section 5 we present a Full Adder (FA) SW TL gate design, validate it by means of micro-magnetic simulations, and compare it with a $MAJ3$ based counterpart. We conclude the paper with some final remarks and future work directions. 



\section{Threshold Logic}

\begin{figure}[htbp]
\centerline{\includegraphics[scale=1.2]{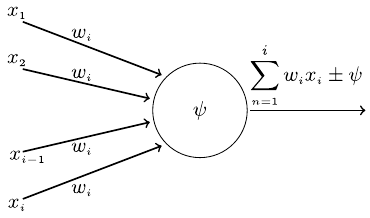}}
\caption{Basic threshold logic gate. 
}
\label{ThresholdLogic}
\end{figure}

As in this paper we bring Threshold rather than Boolean Logic into the framework of SW computing we briefly introduce Threshold Logic (TL) fundamentals. For more inside on this paradigm we refer the reader to \cite{ThresholdLogic}. TL makes use of the basic gate, depicted in Figure \ref{ThresholdLogic}, which evaluates the weighted sum of its inputs, and compares this value with a given threshold value. Note that such a  gate corresponds to the Boolean output neuron introduced in the McCulloch-Pitts neural model \cite{ThresholdLogic} with no learning features. It initially computes (\ref{thresholdbasic}) where $x_{i} \epsilon \{0,1\}$, $w_{i}$ are integer weights, and $\psi$ the threshold value, 

\begin{equation}
    f(x)=\sum_{n=1}^{i}w_{i}x_{i}-\psi\label{thresholdbasic}
\end{equation}

and the actual gate output is computed as: 

$$
    F(x) = sgn(f(x))= \begin{cases} 
                         1, & f(x) \geq 0 \\
                    0, & f(x) < 0
                        \end{cases}\label{sgn}
$$

 Such a TL gate (TLG) can evaluate basic Boolean functions as AND/NAND and OR/NOR (see \ref{ThresholdAND} for TL evaluation of AND), but can also perform more complex calculations. 
  \begin{equation}
    AND = sgn(x_{1} + x_{2}-2)\label{ThresholdAND}
\end{equation}
For example the Full Adder (FA) TL implementation (FA is an essential building block for data processing hardware) can be done with $2$ TLGs \cite{ThreshoildAdder} as follows:
\begin{eqnarray}
    C_{out} &=& sgn(x_{1} + x_{2} + C_{in} - 2) \label{eq:Cout} \\
   Sum &=& sgn(x_{1} + x_{2} + C_{in} - 2C_{out} - 1 ) \label{eq:Sum}
\end{eqnarray}
Previous research demonstrated that TL implementations of basic arithmetic functions can outperform Boolean counterparts in terms of circuit complexity  \cite{Threshold21, Threshold72, ThresholdPeriodic} 
 and TLG implementations have been proposed in CMOS \cite{LatchThresholdGate, CMOSThresholdGate} and in emerging technologies  \cite{SETThresholdGate}.

\section{SW Threshold gate concept}
\begin{figure}[htbp]
    \centering
    \includegraphics[scale=0.3]{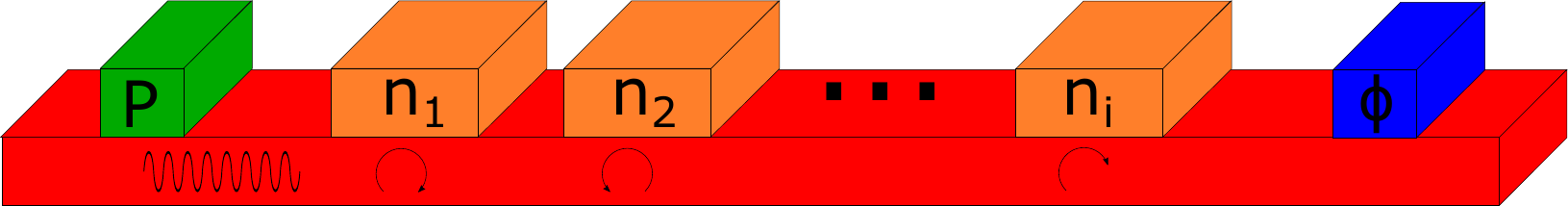}
    \caption{SW phase shift based threshold logic gate.}
    \label{GeneralDevice3D}
\end{figure}

Figure \ref{GeneralDevice3D} depicts our proposed gate concept. The green transducer, which can be an RF antenna or a magneto-electric cell \cite{MEcell}, generates a SW, the orange transducers can only induce a phase shift $\pm p_i$ on the SW, and the blue transducer reads the final net phase change induced by the joint action of the orange transducers. If we use the initial SW phase as reference and phase-shifter $i, i=1\ldots,n$ produce a phase shift proportional with $x_i w_i$, and phase-shifter $n+1$ a phase shift proportional with $\psi$ the net phase shift is proportional with the $f(x)$ value computed by (\ref{thresholdbasic}). Finally, the $TLG$  output according to (\ref{sgn}) is determined by checking the sign of net phase change, i.e., $ \ge 0$ corresponds to logic $1$, and logic $0$ otherwise.   


Thus, to implement an $n$-input TL gate we need $n$ phase-shifters, each of them enabled by $x_i, i=1,\ldots,n$ and producing a phase shift that modulates $w_i$ and one always active phase-shifter inducing a phase shift that modulates $\psi$. Note that for proper operation the actual phase change value per each shifter should be determined in such a way that the net phase shift does not exceed $360^{\circ}$, as further discussed in the following section.

\begin{figure}[htbp]
    \centerline{\includegraphics[scale=0.6]{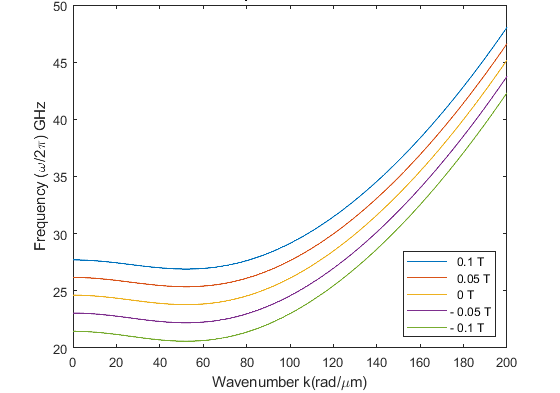}}
    \caption{Dispersion relation (\ref{DispersionRelation}) plot for a \SI{200}{nm} wide and \SI{9}{nm} thick CoFeB waveguide using the parameters specified in \cite{IntroSpinComp}. }
    \label{DispersionRelationMagPlot}
\end{figure}

\begin{figure}[htbp]
    \centerline{\includegraphics[scale=0.3]{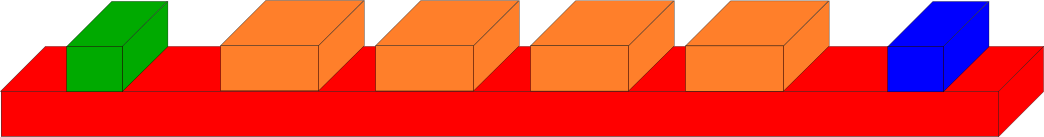}}
    \caption{$FA$ $C_{out}$ evaluation TL gate structure.}
    \label{SumDevice}
\end{figure}

\section{SW phase manipulation}
The behaviour of a SW within a waveguide is captured by the dispersion relation, which reflects the relation between SW frequency and its wave-number or wavelength. According to \cite{Chumak2019FundamentalsOM}  (\ref{DispersionRelation}) presents an approximation of the SW dispersion relation. Note that in our investigation we assume backwards volume SW, which are  waves that travel along the magnetization direction, and only consider waves traveling along the long waveguide axis, i.e., $\theta_k=0$ and $\theta_m=0$. 

\begin{equation}
    \resizebox{.8\columnwidth}{!}{$\omega(k)=\sqrt{({l}{\omega_{_{H}}+\omega_{_{M}}\lambda_{_{ex}}k_{tot})(\,\omega_{_{H}}+\omega_{_{M}}\lambda_{_{ex}}k_{tot}+\omega_{_{M}}F)}}$}, \label{DispersionRelation}
\end{equation}

\noindent where $\omega_{H}=\gamma\mu_{0}H_{eff}$, $\omega_{M}=\gamma\mu_{0}M_{s}$, $\gamma=1.76*10^{11}$rad/(s T) is the gyromagnetic ratio, $\mu_{0}H_{eff}$ is the effective internal magnetic field, $M_{s}$ is the saturation magnetization, $k_{tot}=k^{2}+(n\pi/w)^{2}$, $A_{ex}$ is the exchange constant, and $F$ is expressed as
\begin{equation}
    \resizebox{.9\hsize}{!}{$F=1-gcos^{2}(\theta_{_{k}}-\theta_{_{M}})+\frac{\omega_{_{M}}g(1-g)sin^{2}(\theta_{_{k}}-\theta_{_{M}})}{(\omega_{_{H}}+\omega_{_{M}}\lambda_{ex}\Bigr[{k}^{2}+\Big({n}\pi/w\Big)^{2}\Bigr]}$},\label{FEquation}
\end{equation}
where $\theta_{k}=atan[n\pi/(kw)]$, $g=1-[1-exp(-d\sqrt{k_{tot}})]/(d\sqrt{k_{tot}})$, and $\theta_{k}$ is the angle between SW wave vector and the long axis of the waveguide. Similarly $\theta_{M}$ is the angle between the waveguide magnetization and the waveguide long axis. Figure \ref{DispersionRelationMagPlot} depicts a  dispersion relation (\ref{DispersionRelation}) plot for  external magnetic fields varying from \SI{0.1}{T} to \SI{-0.1}{T}.

As indicated by Maxwell's equations, when a current passes through a wire, it generates a static magnetic field around it. That magnetic field acts on the waveguide and changes the dispersion relation in the area on which the magnetic field is applied. In the dispersion relation $\omega_H$ is related to the effective internal magnetization, which in turn relates to the magnetization acting on the waveguide. Thus by changing the current through the phase-shifter we change the generated magnetic field which correlates to the applied phase shift. The effect is visualised in Figure \ref{DispersionRelationMagPlot} where it can be observed that the dispersion relation moves to higher frequencies for the same wavenumber when applying a higher field and the opposite happens when applying a negative magnetic field by changing the direction of the current flowing through the phase-shifter.  

\begin{figure}
\centerline{\includegraphics[scale=0.6]{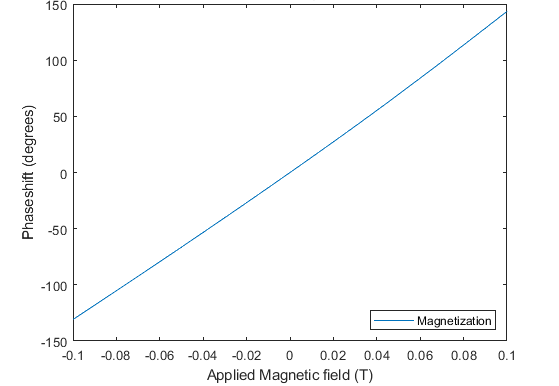}}
\caption{SW phase shift induced by a single \SI{200}{nm} wide phase-shifter when making use of the parameters in  \cite{IntroSpinComp}. 
The relation is linear around no applied field point and becomes less linear for larger field values.}
\label{PhaseShiftRef}
\end{figure}

\begin{figure}
\centerline{\includegraphics[scale=0.6]{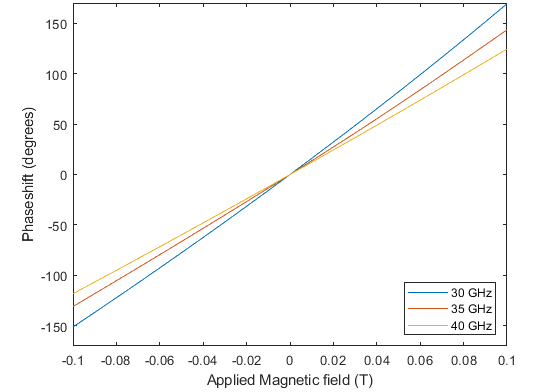}}
\caption{SW phase shift dependency on SW frequency for a single \SI{200}{nm} wide phase-shifter when making use of the parameters  in  \cite{IntroSpinComp}. 
Note that low frequency SWs experience a larger phase shift for the same magnetic field value.}
\label{PhaseShiftDiffFreq}
\end{figure}

\begin{figure}
\centerline{\includegraphics[scale=0.6]{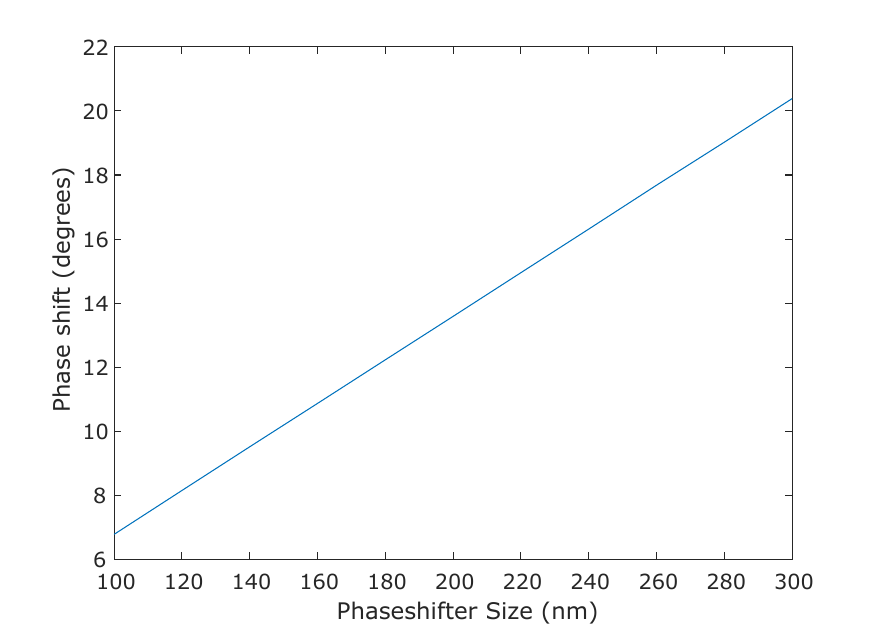}}
\caption{SW phase shift value vs phase-shifter size. 
}
\label{PhaseShiftDiffSize}
\end{figure}

\begin{figure*}[htp]
    \centerline{\includegraphics{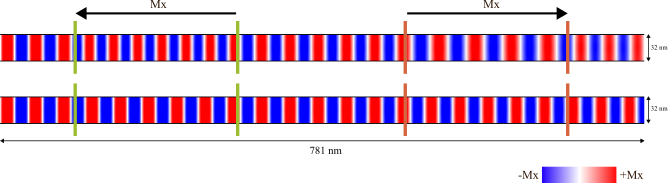}}
    \caption{Working example of our device visually showing the shortening/lengthening of the wavelength under the phase-shifters.}
    \label{VisWaveguideWave}
\end{figure*}

All simulations are performed using mumax3 \cite{Mumax3}. We use a CoFeB waveguide with the parameters from \cite{IntroSpinComp} with dimensions of \SI{2024}{nm} by \SI{32}{nm} by \SI{9}{nm} with a cell size of \SI{2}{nm} by \SI{2}{nm} by \SI{3}{nm}. The material parameters are: $Ms = 1.36*10^{6} A/m$, $Aex = 18.6*10^{-12} J/m$, $Alpha = 0.004$. SW are generated on the left side of the waveguide with a sinusoidal external magnetic field applied on the $x$-axis with a frequency of \SI{35}{GHz} and a field strength of \SI{5}{mT}. The phase-shifters are simulated by applying static magnetic fields in small strips on the waveguide. We initially consider a strip size of \SI{200}{nm}, and when simulating a single phase-shifter we place it in the waveguide center, while when considering multiple phase-shifters they were placed at least \SI{200}{nm} apart from each other to make sure that they did not influence one another. Figure \ref{VisWaveguideWave} depicts a zoom-in capturing the simulated behaviour of a SW travelling through a waveguide  with two phase-shifters covering the waveguide area between the green and red vertical lines, respectively. The green phase-shifter induces a magnetic field in the opposite direction of the waveguide magnetization. When entering this area the SW shrinks because as indicated in Figure \ref{DispersionRelationMagPlot}, a negative magnetic fields shifts the dispersion relation down, which results in larger wavenumber and smaller wavelength for the same SW frequency. A smaller wavelength and same frequency means that the SW travels slower while underneath the phase-shifter resulting in a backward phase shift when comparing with the unperturbed SW in the bottom waveguide. In this simulation, the first phase-shifter produces a $360^{\circ}$ phase shift, which results in an unchanged phase at its output. The red phase-shifter applies a magnetic field along the waveguide magnetization direction, thus it shifts the dispersion relation up, which   results in an increased wavelength and by implication an increase SW  velocity. When exiting the red phase-shifter the wave is significantly forward shifted, as one can clearly observe in the Figure.

To be able to design shifters able to induce phase changes proportional with $TLG$ $w_i$ and $\psi$ values we need to capture the relation between shifter parameters and the induced the phase shift. First, using a single phase-shifter, we simulated the effect of magnetic field strengths on the achieved phase shift angle and the results are presented in Figure \ref{PhaseShiftRef}. When applying small magnetic fields ($\pm\SI{10}{mT}$) the phase shift changes linearly as indicated by $R^2=1.000$, $R^2$ closer to one portraying a higher linearity, rounded to $3$ digits after the comma. When applying larger magnetic fields ($\pm\SI{100}{mT}$) the phase shift becomes less linear ($R^2=0.9993$). Moreover, the system exhibits asymmetry, as when applying large positive magnetic fields, the phase shift magnitude is larger then the one induced by the same magnetic fields applied in the opposite direction. This is related to the dispersion relation non linearity at that frequency, clearly observable in the dispersion relation plot in Figure \ref{DispersionRelationMagPlot}. It is also clear that by substantially increasing the applied magnetic field, we can phase shift with larger angles. 

Next we looked at the behaviour of different frequency SWs when passing under the same phase phase-shifter. The results are visualised in Figure \ref{PhaseShiftDiffFreq} for \SI{30}{GHz}, \SI{35}{GHz}, and \SI{40}{GHz} SWs. As expected from the dispersion relation in Figure \ref{DispersionRelationMagPlot}, the phase shift is larger for lower frequency SWs and smaller for higher frequency SWs. The linearity of the phase shift also decreases for lower frequencies  because the dispersion relation is less linear at those lower frequencies. This can be seen in $R_{30}^2 = 0.9991$, $R_{35}^2 = 0.9993$, and $R_{40}^2 = 0.9998$, which increases when increasing the SW frequency. This can be also intuitively deduce by observing the  dispersion relation plot: as we go to a lower frequency, the slope becomes smaller, meaning that that the reverse of the slope becomes bigger resulting in larger positive and negative phase shifts.

Following up, when decreasing the phase-shifter size we observe a phase shift decrease, while a size increase results in a larger phase shift. Figure \ref{PhaseShiftDiffSize} depicts the  due to an external field of \SI{0.01}{T} phase shift dynamics when increasing the  phase-shifter size from \SI{100}{nm} to \SI{300}{nm}. As expected, a larger phase-shifter induces a larger phase shift as the SW is slowed or sped up over a longer propagation distance.

\section{Full Adder design and simulation}

\begin{figure}[htp]
    \centerline{\includegraphics[scale =1.1]{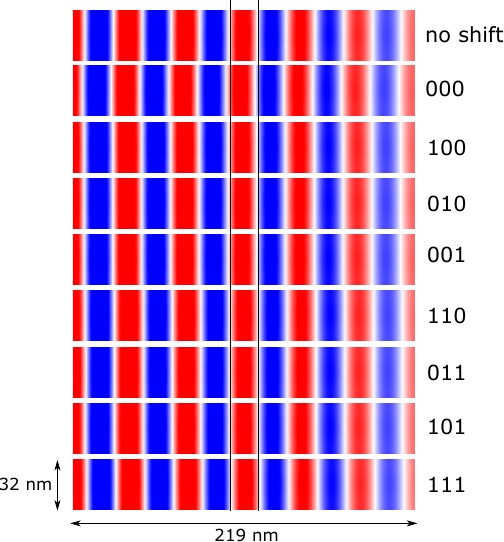}}
    \caption{
    $C_{out}$ $TLG$ output phase shift for all possible $FA$ input combinations.
    }
    \label{VisWaveguideWaveCout}
\end{figure}

\begin{figure}[htp]
    \centerline{\includegraphics[scale=1.1]{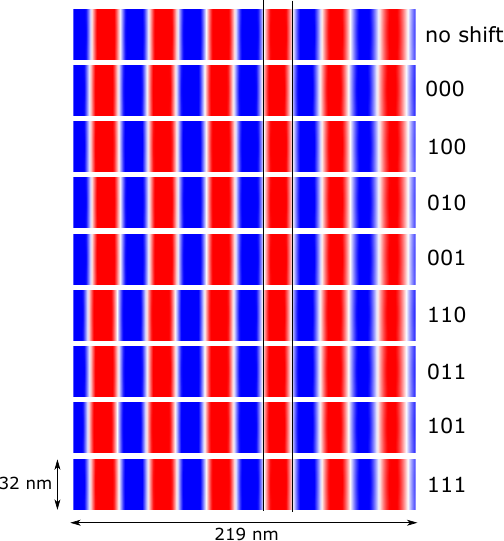}}
    \caption{
    $Sum$ $TLG$ output phase shift for all possible $FA$ input combinations.
    }
    \label{VisWaveguideWaveSum}
\end{figure}

In this section we present the TL implementation of the Full Adder outputs $C_{out}$ and $Sum$ by means of two TL gates each of them including $4$ and $5$ phase-shifters, respectively. We made use of  \SI{100}{nm} wide phase-shifters calibrated to induce a $10^{\circ}$ phase shift  per input weight unit. This corresponds to an applied external field of  $w_i~\times$~\SI{0.0147}{T} when the input $x_i = 1$ and no field otherwise, on each $TLG$ input.  Table \ref{TableCout} and Table \ref{TableSum} present the net phase shift observed by means of micro-magnetic simulations at the output of the $TLG$ producing $C_{out}$ and $Sum$, respectively, for all possible FA input combinations. Given that when reading out a phase shift of $0^{\circ}$ or higher, the $TLG$ outputs a logic $1$ and a negative phase shift results in a logic $0$, one can easily observe that the two $FA$ outputs are correctly evaluated, according to the $FA$ truth table. Note that the phase shifts are not exact multiples of $10^{\circ}$ due to the dispersion relation non-symmetry when applying the same field  magnitude in opposite directions. The  phase shift produced by all possible $FA$ input combinations is also visualised in Figure \ref{VisWaveguideWaveCout} and \ref{VisWaveguideWaveSum}, for the $TLG$ producing $C_{out}$ and $Sum$ outputs, respectively. 

\begin{table}
\centering
\caption{Net Phase Shift $\Delta \phi$ for $C_{out}$}
\begin{tabular}{cccS[table-format=2.4]c}
\hline
\textbf{$a$} & \textbf{$b$} & \textbf{$C_{in}$} & \textbf{$\Delta \phi$} & \textbf{$C_{out}$} \\
\hline
0&0&0&-19.6668&0 \\
1&0&0&-9.6686&0 \\
0&1&0&-9.6748&0 \\
0&0&1&-9.7002&0 \\
1&1&0&0.3165&1 \\
0&1&1&0.3260&1 \\
1&0&1&0.3306&1 \\
1&1&1&10.3377&1 \\
\hline
\end{tabular}\label{TableCout}
\end{table}

\begin{table}
\centering
\caption{Net Phase Shift $\Delta \phi$ for $Sum$}
\begin{tabular}{ccccS[table-format=-1.4]c}
\hline
\textbf{$a$} & \textbf{$b$} & \textbf{$C_{in}$} & \textbf{$C_{out}$} & \textbf{$\Delta \phi$} & \textbf{Sum} \\
\hline
0&0&0&0&-9.9041&0 \\
1&0&0&0&0.0987&1 \\
0&1&0&0&0.1043&1 \\
0&0&1&0&0.0891&1 \\
1&1&0&1&-9.5451&0 \\
0&1&1&1&-9.5632&0 \\
1&0&1&1&-9.5711&0 \\
1&1&1&1&0.4538&1 \\
\hline
\end{tabular}\label{TableSum}
\end{table}

To get inside into the potential practical implications of the proposed SW $TLG$ we assume as discussion vehicle the SW implementation of a Full Adder (FA), which is a heavily utilized basic building block in computation platform designs, and conceptually compare $MAJ3$ and $TLG$ based implementations. 

The $MAJ3$ FA implementation relies on the following equations \cite{MajorityAdder}:
\begin{eqnarray}
    C_{out} &=& MAJ3(x_1,x_2,C_{in}) \\
    Sum &=& MAJ3(\overline{C_{out}}, MAJ3(x_1,x_2, \overline{C_{out}}), C_{in}).
\end{eqnarray}
Thus, it requires $3$ $MAJ3$ gates and exhibit a $2$ $MAJ3$ gates delay, actually a bit larger  as while generating $\overline{C_{out}}$ does not require an inverter it induces a small delay overhead due to the extra output transducer. 
The $TLG$ FA implementation relies on (\ref{eq:Cout}) and (\ref{eq:Sum}) and requires $2$ $TLG$ gates and exhibits a $2$ $TLG$ gates delay, thus it clearly outperforms the $MAJ3$ implementation in terms of area. 

While a more accurate comparison requires a detailed design of the two $FA$ implementation and it is subject to future work we can also have a glimpse into other aspects that may make $TLG$s more attractive. A $MAJ3$ gate requires $3$ transducers to generate the input SWs and one to read the output, thus the entire $FA$ requires $12$ transducers (actually it may need one more for reading $\overline{C_{out}}$). As an $n$-input $TLG$ requires a transducer to generate the SW, $n+1$ shifters, and one output reading transducer, the $TLG$ $FA$ requires $4$ transducers and $9$ phase-shifters. Given that phase-shifters are potentially smaller  than transducers the $TLG$ waveguides are smaller, thus faster. Moreover, while transducers are RF operated the shifters require DC inputs, which may induce further advantages in terms of power consumption. Last, but not least
inline majority gates may  significantly suffer from imprecise manufacturing as the  transducers should be placed at precise distances from each other in order to enable proper $MAJ3$ gate operation, while $TLG$s are more robust as the phase-shifters positioning does not need to be that precise.


\section{Conclusions}
In this paper we introduced a novel Threshold Logic (TL) inspired Spin Wave (SW) based computing paradigm, which relies on successive SW phase rotations applied to one single SW instead of on SW interference. We  introduced the SW TL gate  concept and discussed ways to mirror TL gate weight and threshold values into physical device  parameters. We designed and demonstrated proper operation of an SW TL based Full Adder (FA) by means of micro-magnetic simulations. We also provided high level evidence that our proposal can potentially outperform functionally equivalent SW interference based implemented counterparts. 

\section{acknowledgements}
   This project is supported by the imec industrial affiliate program on Beyond CMOS Logic. It has also been supported by SPIDER, EC contract number 101070417; Topic advanced spintronics; Unleashing spin into the next generation ICs (RIA).

\bibliographystyle{abbrv}
\bibliography{bibliography}

\begin{thebibliography}{10}

\bibitem{NanoMagnet}
S.~D. Bader.
\newblock Colloquium: Opportunities in nanomagnetism.
\newblock {\em Rev. Mod. Phys.}, 78:1--15, Jan 2006.

\bibitem{GrapheneComp}
S.~K. Banerjee, L.~F. Register, E.~Tutuc, D.~Basu, S.~Kim, D.~Reddy, and A.~H. MacDonald.
\newblock Graphene for cmos and beyond cmos applications.
\newblock {\em Proceedings of the IEEE}, 98(12):2032--2046, 2010.

\bibitem{MEcell}
S.~Cherepov, P.~Khalili~Amiri, J.~G. Alzate, K.~Wong, M.~Lewis, P.~Upadhyaya, J.~Nath, M.~Bao, A.~Bur, T.~Wu, G.~P. Carman, A.~Khitun, and K.~L. Wang.
\newblock {Electric-field-induced spin wave generation using multiferroic magnetoelectric cells}.
\newblock {\em Applied Physics Letters}, 104(8):082403, 02 2014.

\bibitem{Chumak2019FundamentalsOM}
A.~V. Chumak.
\newblock Fundamentals of magnon-based computing.
\newblock {\em arXiv: Mesoscale and Nanoscale Physics}, 2019.

\bibitem{ThresholdPeriodic}
S.~Cotofana and S.~Vassiliadis.
\newblock Periodic symmetric functions, serial addition, and multiplication with neural networks.
\newblock {\em IEEE Transactions on Neural Networks}, 9(6):1118--1128, 1998.

\bibitem{FinFetScaling}
U.~K. Das and T.~K. Bhattacharyya.
\newblock Opportunities in device scaling for 3-nm node and beyond: Finfet versus gaa-fet versus ufet.
\newblock {\em IEEE Transactions on Electron Devices}, 67(6):2633--2638, 2020.

\bibitem{EnergyCons}
C.~Y. Gelenbe~E.
\newblock The impact of information technology on energy consumption and carbon emissions.
\newblock {\em ACM}, pages 1--15, 2015.

\bibitem{ReviewSpin}
A.~Hirohata, K.~Yamada, Y.~Nakatani, I.-L. Prejbeanu, B.~Diény, P.~Pirro, and B.~Hillebrands.
\newblock Review on spintronics: Principles and device applications.
\newblock {\em Journal of Magnetism and Magnetic Materials}, 509:166711, 2020.

\bibitem{SpintronicsReview}
A.~Hirohata, K.~Yamada, Y.~Nakatani, I.-L. Prejbeanu, B.~Diény, P.~Pirro, and B.~Hillebrands.
\newblock Review on spintronics: {Principles} and device applications.
\newblock {\em Journal of Magnetism and Magnetic Materials}, 509:166711, Sept. 2020.

\bibitem{Quantum}
J.~Hutchby, G.~Bourianoff, V.~Zhirnov, and J.~Brewer.
\newblock Extending the road beyond cmos.
\newblock {\em IEEE Circuits and Devices Magazine}, 18(2):28--41, 2002.

\bibitem{GNRpaper}
Y.~Jiang, N.~C. Laurenciu, H.~Wang, and S.~D. Cotofana.
\newblock Graphene nanoribbon based complementary logic gates and circuits.
\newblock {\em IEEE Transactions on Nanotechnology}, 18:287--298, 2019.

\bibitem{SETThresholdGate}
C.~Lageweg, S.~Cotofana, and S.~Vassiliadis.
\newblock A linear threshold gate implementation in single electron technology.
\newblock In {\em Proceedings IEEE Computer Society Workshop on VLSI 2001. Emerging Technologies for VLSI Systems}, pages 93--98, 2001.

\bibitem{ThreshoildAdder}
C.~Lageweg, S.~Cotofana, and S.~Vassiliadis.
\newblock A full adder implementation using set based linear threshold gates.
\newblock In {\em 9th International Conference on Electronics, Circuits and Systems}, volume~2, pages 665--668 vol.2, 2002.

\bibitem{IntroSpinComp}
A.~Mahmoud, F.~Ciubotaru, F.~Vanderveken, A.~V. Chumak, S.~Hamdioui, C.~Adelmann, and S.~Cotofana.
\newblock Introduction to spin wave computing.
\newblock {\em Journal of Applied Physics}, 128(16):161101, 2020.

\bibitem{SWFanout}
A.~Mahmoud, F.~Vanderveken, C.~Adelmann, F.~Ciubotaru, S.~Hamdioui, and S.~Cotofana.
\newblock {Fan-out enabled spin wave majority gate}.
\newblock {\em AIP Advances}, 10(3):035119, 03 2020.

\bibitem{SpinwaveMajority}
A.~Mahmoud, F.~Vanderveken, F.~Ciubotaru, C.~Adelmann, S.~Cotofana, and S.~Hamdioui.
\newblock n-bit data parallel spin wave logic gate.
\newblock In {\em 2020 Design, Automation \& Test in Europe Conference \& Exhibition (DATE)}, pages 642--645, 2020.

\bibitem{ThresholdLogic}
S.~Muroga.
\newblock Threshold logic and its applications.
\newblock 1971.

\bibitem{LatchThresholdGate}
M.~Padure, S.~Cotofana, C.~Dan, M.~Bodea, and S.~Vassiliadis.
\newblock A new latch-based threshold logic family.
\newblock In {\em 2001 International Semiconductor Conference. CAS 2001 Proceedings (Cat. No.01TH8547)}, volume~2, pages 531--534 vol.2, 2001.

\bibitem{CMOSThresholdGate}
M.~Padure, S.~Cotofana, S.~Vassiliadis, C.~Dan, and M.~Bodea.
\newblock A low-power threshold logic family.
\newblock In {\em 9th International Conference on Electronics, Circuits and Systems}, volume~2, pages 657--660 vol.2, 2002.

\bibitem{MajorityAdder}
V.~Pudi, K.~Sridharan, and F.~Lombardi.
\newblock Majority logic formulations for parallel adder designs at reduced delay and circuit complexity.
\newblock {\em IEEE Transactions on Computers}, 66(10):1824--1830, 2017.

\bibitem{BigData}
M.~E. Richard L.~Villars, Carl W.~Olofson.
\newblock Big data: what it is and why you should care.
\newblock {\em IDC}, pages 1--14, 2011.

\bibitem{InlineMajorityGate}
G.~Talmelli, T.~Devolder, N.~Träger, J.~Förster, S.~Wintz, M.~Weigand, H.~Stoll, M.~Heyns, G.~Schütz, I.~P. Radu, J.~Gräfe, F.~Ciubotaru, and C.~Adelmann.
\newblock Reconfigurable submicrometer spin-wave majority gate with electrical transducers.
\newblock {\em Science Advances}, 6(51):eabb4042, 2020.

\bibitem{CMOSscaling}
S.~Thompson, R.~Chau, T.~Ghani, K.~Mistry, S.~Tyagi, and M.~Bohr.
\newblock In search of "forever," continued transistor scaling one new material at a time.
\newblock {\em IEEE Transactions on Semiconductor Manufacturing}, 18(1):26--36, 2005.

\bibitem{Mumax3}
A.~Vansteenkiste, J.~Leliaert, M.~Dvornik, M.~Helsen, F.~Garcia-Sanchez, and B.~Van~Waeyenberge.
\newblock The design and verification of mumax3.
\newblock {\em AIP Advances}, 4(10):107133, 2014.

\bibitem{Threshold72}
S.~Vassiliadis and S.~Cotofana.
\newblock 7|2 counters and multiplication with threshold logic.
\newblock In {\em Conference Record of The Thirtieth Asilomar Conference on Signals, Systems and Computers}, volume~1, pages 192--196 vol.1, 1996.

\bibitem{Threshold21}
S.~Vassilladis, S.~Cotofana, and K.~Bertels.
\newblock 2-1 addition and related arithmetic operations with threshold logic.
\newblock {\em IEEE Transactions on Computers}, 45(9):1062--1067, 1996.

\bibitem{SpintronicsNeuromorph}
D.~Zhang, W.~Zhao, L.~Zeng, K.~Cao, M.~Wang, S.~Peng, Y.~Zhang, Y.~Zhang, J.-O. Klein, and Y.~Wang.
\newblock All {Spin} {Artificial} {Neural} {Networks} {Based} on {Compound} {Spintronic} {Synapse} and {Neuron}.
\newblock {\em IEEE Transactions on Biomedical Circuits and Systems}, 10(4):828--836, Aug. 2016.

\end{thebibliography}

\appendix

\end{document}